\documentclass[a4paper,english]{article}
\usepackage{fullpage}
\usepackage{color}
\usepackage{graphicx}
\usepackage[colorinlistoftodos]{todonotes}
\usepackage{booktabs}
\usepackage{multirow}
\usepackage{amsmath,amssymb}

\usepackage[affil-it]{authblk}  

\newtheorem{theorem}{Theorem}
\newtheorem{proposition}{Proposition}

\newtheorem{lemma}{Lemma}

\newtheorem{definition}{Definition}

\newenvironment{proof}{\medskip\noindent\textbf{Proof.}}{{}\hfill$\Box$\\}

\let\eps=\epsilon
\newcommand{\costlp}{\mathit{COST_{\mathrm{LP}}}}
\newcommand{\T}{\mathcal T}
\renewcommand{\S}{\mathcal S}
\newcommand{\E}{\mathbf E}
\newcommand{\poly}{\mathrm{poly}}
\newcommand{\hop}[1]{$\ensuremath #1$-hopsets}

\DeclareMathOperator{\argmin}{argmin}
\DeclareMathOperator{\argmax}{argmax}
\DeclareMathOperator{\cut}{\mathit{Cut}}
\DeclareMathOperator{\reach}{\mathit{Reach}}
\DeclareMathOperator{\width}{\mathit{Width}}
\DeclareMathOperator{\sk}{\mathit{k}}
\newcommand{\G}{{\wtilde{G}}}
\newcommand{\Tree}{{\wtilde{T}}}
\newcommand{\wtilde}[1]{\widetilde{#1}}

\makeatletter
\def\blfootnote{\gdef\@thefnmark{}\@footnotetext}
\makeatother

\title{Exploiting Hopsets: Improved Distance Oracles for Graphs of Constant Highway Dimension and Beyond}

\author[1]{Siddharth Gupta}
\author[2]{Adrian Kosowski}
\author[2]{Laurent Viennot}

\affil[1]{Ben-Gurion University of the Negev, Israel\footnote{Supported in part by the Zuckerman STEM Leadership Program.}}
\affil[2]{Inria, Paris University, France\footnote{Supported by Irif CNRS laboratory and ANR projects DESCARTES (ANR-16-CE40-0023) and DISTANCIA (ANR-17-CE40-0015).}}

\begin{document}
\maketitle

\begin{abstract}
  For fixed $h \geq 2$, we consider the task of adding to a graph $G$ a set of weighted shortcut edges on the same vertex set, such that the length of a shortest $h$-hop path between any pair of vertices in the augmented graph is exactly the same as the original distance between these vertices in $G$. A set of shortcut edges with this property is called an \emph{exact $h$-hopset} and may be applied in processing distance queries on graph $G$. In particular, a $2$-hopset directly corresponds to a distributed distance oracle known as a \emph{hub labeling}. In this work, we explore centralized distance oracles based on $3$-hopsets and display their advantages in several practical scenarios. In particular, for graphs of constant highway dimension, and more generally for graphs of constant skeleton dimension, we show that $3$-hopsets require \emph{exponentially} fewer shortcuts per node than any previously described distance oracle,
  and also offer a speedup in query time when compared to simple oracles based on a direct application of $2$-hopsets. Finally, we consider the problem of computing minimum-size $h$-hopset (for any $h \geq 2$) for a given graph $G$, showing a polylogarithmic-factor approximation for the case of unique shortest path graphs. When $h=3$, for a given bound on the space used by the distance oracle, we provide a construction of hopset achieving polylog approximation both for space and query time compared to the optimal $3$-hopset oracle given the space bound.

\end{abstract}


\section{Introduction}
An exact $h$-hopset for a weighted graph $G$ is a weighted edge set, whose addition to the graph guarantees that every pair of vertices has a path between them with at most $h$ edges (\emph{hops}) and whose length is exactly the length of shortest path between the vertices. 

The concept of a hopset was first explicitly described by Cohen~\cite{Cohen00} in its \emph{approximate} setting, in which the length of $h$-hop path between a pair of vertices in the hopset should approximate the length of the shortest path in $G$. Hopsets were introduced in the context of parallel computation of approximate shortest paths. In this paper, we study hopsets in their exact version, with the general objective of optimizing exact shortest path queries. 

Data structures which allow for querying distance between any pair of vertices of a graph have been intensively studied under the name of \emph{distance oracles}. The efficiency of an exact distance oracle is typically measured by the interplay between the \emph{space} requirement of the representation of the data structure and its \emph{decoding time}. It is a well-established empirical fact that many real-world networks admit efficient (i.e., low-space and fast) distance oracles~\cite{DBLP:conf/www/AkibaIY14,DBLP:conf/esa/DellingGPW14}. 
A key example here concerns transportation networks, and specifically road networks, which are empirically known~\cite{reach,RE,hw10} to be augmentable by carefully tailored sets of shortcut edges, allowing for shortest-path computation. These sets of shortcuts may be hopsets (as is the case for the hub-labeling approach which effectively implements a $2$-hopset), but may also be considered in some related (and frequently more involved) framework, such as contraction hierarchies~\cite{DBLP:conf/wea/GeisbergerSSD08} or transit-node routing~\cite{DBLP:conf/dimacs/BastFM06}. 

An interesting theoretical insight due to Abraham et al.~\cite{DBLP:journals/jacm/AbrahamDFGW16,MS-TR,hw10} provides theoretical bounds on the number of shortcuts required in all of the above-mentioned frameworks. They introduce a parameter describing the structure of shortest paths within ball neighborhoods of a graph, called \emph{highway dimension} $\tilde h$. They also express the number of shortcuts that need to be added for each node so as to achieve shortest-path queries in a graph of $n$ nodes with weighted diameter $D$ as a polynomial of $\tilde h$, $\log n$, and $\log D$; this approach has been extended in subsequent work~\cite{hwVC,KV-SODA2017}. The value of $\tilde h$ is known to be small in practice (e.g., typically $\tilde h < 100$ for continental-sized road networks~\cite{MS-TR}), and does indeed appear to be inherently linked to the size of the required shortcut sets. In fact, empirical tests have suggested that the (average) number of necessary shortcuts per node is in fact very close to $\tilde h$, laying open the question of whether the additional dependence of the number of shortcuts on logarithmic factors in  $n$ and $D$ may be an artifact of the theoretical analysis of the oracles, which for each node require a separate shortcut for every ``scale'' of distance.

\subsection{Results and Organization of the Paper}

Our main result is to provide strong evidence that the dependence of the number of shortcuts on such logarithmic factors in $n$ and $D$ is indeed not essential, and we design a simple distance oracle based on a $3$-hopset in which the number of shortcuts per node depends only on $\tilde h$, $\log \log n$, and the logarithm of the average edge length. This result is in fact shown in the framework of a strictly broader class of graphs, namely, graphs with a bounded value of a parameter known as \emph{skeleton dimension $k$} ($k \leq \tilde h$), describing the width of the shortest-path tree of a node after pruning all branches at a constant fraction $\alpha$ of their depth. Considering various ranges of fraction $\alpha$ for increasing distance ranges was a novel key step for improving over~\cite{KV-SODA2017,KV-long2016} from a 2-hopset construction to a 3-hopset construction.

\begin{table}
\scalebox{0.9}{
 \begin{tabular}{lcccc}
\toprule
\multirow{2}{*}{Distance oracle} & \multicolumn{2}{c}{Treewidth $t$} & \multicolumn{2}{c}{Skeleton dimension $k$}\\
 & \emph{Size} & \emph{Time} & \emph{Size} & \emph{Time}\\
\midrule
$2$-hopset (hubs): & $n\cdot O(t\log n)$ & $O(t + \log \log n)$ & $n\cdot O(k\log n)$ & $O(k \log n)$ \\ 
\textbf{$3$-hopset}: & $n\cdot O(t\log \log n)$ & $O(t^2 \log^2 \log n)$ & $n\cdot O(k \log k \log \log n)$ & $O(k^2 \log^2 k \log^2 \log n)$\\
\bottomrule\vspace{-1mm}
\end{tabular}}
\caption{Comparison of distance oracles based on $2$-hopsets (hub labeling~\cite{Cohen00,Gavoille:2004:DLG:1036161.1036165,KV-SODA2017}) and $3$-hopsets (this paper). Size represents the number of shortcut edges in the hopset, i.e., the number of $O(\log n)$-bitsize words when measuring oracle size. The main results concern skeleton dimension and are stated in simplified form, assuming average edge length at most $O(\poly\log n)$, with expected query times given for both types of oracles.\vspace{-3mm}}\label{tab1}
\end{table}

From a general perspective, our connection between $h$-hopsets and distance oracles is original and offers new perspectives for studying the trade-off between size and query time of distance oracles. To exemplify this, we provide a construction of $h$-hopsets for graphs of treewidth $t$ following a classical approach in  pre-processing product queries on trees~\cite{AlonSchieber,Chazelle87a}. For $3$-hopsets, we obtain a distance oracle with quadratic dependency in $t$ which improves over the construction of \cite{ChaudhuriZ00} (which has cubic dependency) for $t=\omega(\log^2\log n)$.
The space and time-bounds of oracles based on $3$-hopsets are presented in Table~\ref{tab1}, and compared with the corresponding parameters of oracles based on $2$-hopsets. For the case of constant skeleton dimension or constant treewidth, we remark that using a $3$-hopset instead of a $2$-hopset reduces the number of shortcuts per node from $O(\log n)$ to $O(\log \log n)$ while achieving a query time of $O(\log^2 \log n)$.

A classical assumption (applied, e.g., in almost all literature on transportation networks) resides in the uniqueness of shortest paths. It can be made without loss of generality by slightly perturbing the weights of the edges or by using appropriate tie break rules. In this context of \emph{unique shortest path graph} (USP) graphs where there is a unique shortest path $P_{uv}$ between any two nodes $u$ and $v$, we propose an LP-based approximation algorithm for constructing $h$-hopsets with size within a polylog factor from optimal. Our construction can be seen as a non-trivial generalization of the prehub labeling introduced in~\cite{AngelidakisMO17} from 2 to more hops. In the case $h=3$, we further extend our approach to provide an algorithm which constructs distance oracles in USP graphs based on $3$-hopsets, with (approximate) optimality guarantees on size and query time. The form of guarantees we obtain is again novel: for a given size bound $\S$ of 3-hopset based oracle, we construct an oracle with size larger than $\S$ by at most a polylog factor which has average query time within a polylog factor of the performance achieved by the best oracle with size $\S$.

The rest of the paper is organized as follows. In Section~\ref{se:prelim}, we introduce the necessary notions related to $h$-hopset and give a general approach for how a $h$-hopset can be used as a distance oracle, focusing on the special case of $h=3$. In Section~\ref{se:skeleton}, we provide our first main result, using $3$-hopsets to obtain improved (smaller and faster) distance oracles in graphs with bounded skeleton dimension. In Section~\ref{se:lp}, we present our second main result about approximating $h$-hopsets and constructing $3$-hopset based oracles in USP graphs. Finally, Section~\ref{se:boundedtw}, we show how to construct efficient $h$-hopsets and 3-hopset based oracles for bounded treewidth graphs.

Our work is presented in the context of weighted undirected graphs, but all results can easily be extended to weighted directed graphs.

\subsection{Other Related Work}

\paragraph{Hopsets.} 
Exact hopsets were implicitly constructed in the context of single-source shortest paths parallel computation~\cite{UY91,KS97,Cohen97,SS99}. Such works study the work versus time trade-offs of such computation.
Cohen~\cite{Cohen00} explicitly introduced the notion of $(h,\epsilon)$-hopset of $G$ as set $H$ of weighted edges such that paths of at most $h$ hops in $G\cup H$ have length within $(1+\eps)$ of the corresponding shortest path in $G$. The parameter $h$ is called the \emph{hopbound}. For any graph $G$ and $\eps,\eps' > 0$, she proposed a construction of $(O(\poly\log n),\eps)$-hopset of $G$ with size $O(n^{1+\eps'})$. More recently, Elkin et al.~\cite{ElkinN16} proposed the construction of $(O(\epsilon^{-1}\log\kappa)^{\log\kappa}, \eps)$-hopset with $O(n^{1+1/\kappa}\log n\log \kappa)$ edges for any $\eps >0$ and integral $\kappa\ge 1$.
Abboud et al.~\cite{ABP17} 
recently showed the optimality of the Elkin et al.~\cite{ElkinN16} result. In particular, they showed that for any $\delta > 0$ and integer $k$, any hopset of size less than $n^{1+\frac{1}{2^{k+1}-1}-\delta}$ must have hop bound $h = \Omega(c_k / \epsilon^{k+1})$, where $c_k$ is a constant depending only on $k$. The linear size case was then improved in~\cite{ElkinN17}.
As far as we know, exact hopsets (with $\eps=0$) have not been explicitly studied. However, they are related to the following well studied notion.

\paragraph{Hopsets vs.\ TC-spanners.} In directed graphs, a hopset can be seen as a special case of an \emph{$h$-transitive-closure spanner ($h$-TC-spanner)}.
Hopsets and TC-spanners are fundamental graph-theoretic objects and are widely used in various settings from distance oracles to pre-processing for range queries in sequential or parallel setting or even in property testing.
The concept of adding transitive arcs to a digraph in order to reduce its diameter was introduced by Thorup~\cite{Thorup95} in the context of parallel processing. Bhattacharyya et al.~\cite{BhattacharyyaGJRW12} defined an $h$-TC-spanner of an unweighted digraph $G$ as a digraph $H$ with same transitive closure as $G$ and diameter at most $h$. They note that this is a  central concept in a long line of work around pre-processing a tree for range queries~\cite{AlonSchieber,Chazelle87a,Thorup97}. A TC-spanner can also be defined as a spanner (for the classical spanner definition~\cite{PelegS89}) of the transitive closure of a graph that has bounded diameter. We will see that an exact $h$-hopset defines a $h$-TC-spanner but that the converse is not necessarily true. Bhattacharyya et al.~\cite{BhattacharyyaGJRW12} proposed a construction of $h$-TC-spanner of size $O(n\log n\lambda_h(n))$ for $H$-minor-free graphs (where $\lambda_h$ denotes the $h$th-row inverse Ackermann function, cf.\ Section~\ref{se:acker}).

\paragraph{Exact Distance Oracles.}
A long line of research studies the interplay between data structure space and query decoding time. A lot of attention has been given to distance oracles for planar graphs~\cite{Djidjev96,Arikati96,Xu2000,Cabello12,FR06,Cohen-AddadDW17,GawrychowskiMWW18}, and it has recently been shown that a distance oracle with $O(n^{1.5})$ space and $O(\log n)$ query-time is possible~\cite{GawrychowskiMWW18}.
In the context of weighted directed graphs with treewidth $t$, Chaudhuri and Zaroliagis~\cite{ChaudhuriZ00} propose a distance oracle using $O(t^2n\lambda_h(n))$ space and $O(t^3 h+\lambda_h(n))$ query time for integral $h>1$ where $\lambda_h$ is the $h$th-row inverse Ackermann function (as defined in Subsection~\ref{se:acker}).
In the context of unweighted graphs with treewidth $t$, Farzan and Kamali~\cite{FarzanK14} obtain distance oracles with $O(t^3\log^3t)$ query time using optimal space (within low order terms). This construction heavily relies on the unweighted setting as exhaustive look-up tables are constructed for handling graphs with polylogarithmic size.

\paragraph{Distance Labelings and 2-Hopsets.}
The distance labeling problem is a special case of a distributed distance oracle, and consists of assigning labels to the nodes of a graph such that the distance between two nodes $s$ and $t$ can be computed from the labels of $s$ and $t$ (see, e.g., \cite{Gavoille:2004:DLG:1036161.1036165}). 

The notion of 2-hopset studied in this work coincides with the special case of two-hop distance labeling (also called \emph{hub-labeling}), where labels are constructed from hub sets: in hub-labeling, a small hub set $S(u)\subseteq V(G)$ is assigned to each node of a graph $G$ such that for any pair $u, v$ of nodes, the intersection of hub sets $S(u) \cap S(v)$ contains a node on a shortest $u-v$ path. Such a construction is formally proposed in~\cite{Cohen:2003:RDQ:942270.944300} and is implicitly introduced by Gavoille et al.~\cite{Gavoille:2004:DLG:1036161.1036165} and applied to graphs of treewidth $t$ with labels of $O(t\log n)$ size and allows to answer distance queries in $O(t\log n)$ time; the hub sets have a hierarchical structure, which allows for an improvement of query time to $O(t\log \log n)$ time by a binary search over levels. Hub labelings are the currently best known distance labelings for sparse graphs, achieving sublinear node label size~\cite{Sublinear,DBLP:conf/wdag/GawrychowskiKU16}, and may also be used to provide a 2-additive-approximation for distance labeling in general graphs using sublinear-space labels~\cite{DBLP:conf/wdag/GawrychowskiKU16}.

In graphs of bounded highway dimension, hub labels were among the first identified distance oracles to provide label size and query time polynomial in the highway dimension and polylogarithmic in other graph parameters~\cite{hw10}. This result was then extended to the more general class of graphs with bounded skeleton dimension~\cite{KV-SODA2017,KV-long2016}. 

Hub sets with near to optimal size can be constructed in polynomial time. A greedy set cover-type $O(\log n)$-approximation algorithm (with respect to average size of a hub set) was proposed by Cohen et al.~\cite{Cohen:2003:RDQ:942270.944300}. For the case of USP graphs, this approximation ratio was improved by Angelidakis et al.~\cite{AngelidakisMO17} to the logarithm of the graph hop-diameter $D_H$, i.e., the maximum number of hops of a shortest path in $G$, showing an approximation gap between USP and non-USP graphs.

\section{Preliminaries}\label{se:prelim}

\subsection{Definitions}
We are given a weighted undirected graph $G=(V,E,\omega)$ where $\omega : E\rightarrow \mathbb{R^+}$ associates a weight with each edge of $G$.
For a positive integer parameter $h$ and a pair $u,v\in V$, the \emph{$h$-limited distance} between $u$ and $v$, denoted $d^h_G(u,v)$, is defined as the length of the shortest path from $u$ to $v$ that contains at most $h$ edges (aka \emph{hops}). The usual shortest path distance can be defined as $d_G(u,v)=d_G^{n-1}(u,v)$. For the sake of brevity, we often let $uv$ denote the pair $\{u,v\}$ representing an edge from $u$ to $v$.

\begin{definition}
An (exact) $h$-hopset for a weighted  graph $G$ is a set of edges $H$ such that $d^h_{G\cup H}(u,v) = d_G(u,v)$ for all $u,v$ in $V(G)$ where $G\cup H=(V,E\cup H,w')$ is the graph augmented with edges of the hopset with weights $w'(u,v)=d_G(u,v)$ for $uv\in H$ and $w'(u,v)=w(u,v)$ for $uv\in E\setminus H$. The parameter $h$ is called the \emph{hopbound} of the hopset. Edges from set $H$ are called \emph{shortcuts} in $G$.
\end{definition}
By convention, we will assume that all self-loops at nodes of $V$ are included in $H$.
Thus, $G\cup H$ is a graph whose $h$-th power in the $(\min,+)$ algebra on $n\times n$ matrices of edge weights corresponds to the transitive closure of the weight matrix of graph $G$. 

Equivalently, a $h$-hopset can be defined as a set $H$ of edges such that for any pair $s,t$, there exists a path $P$ of at most $h$ edges from $s$ to $t$ in $G\cup H$ and a shortest path $Q$ from $s$ to $t$ in $G$ such that all nodes of $P$ belong to $Q$ and appear in the same order. 
Note that a $h$-hopset is completely specified by its set $H$ of edges as the associated weights are deduced from distances in the graph.

\subsection{Using a Hopset as a Distance Oracle}

Hopsets may be used to answer shortest-path queries in a graph $G=(V,E)$. In general, given a hopset $H$, the na\"ive way to approach a query for $d_G(u,v)$ for a given node pair $u,v$ is to perform a bidirectional Dijkstra search in graph $G\cup H$ from this node pair, limited to a maximum of $\lceil h/2 \rceil$ hops distance from each of these nodes. We have, in particular for any pair $u,v \in V$:
$$
d_G(u,v) = \min_{w\in V} (d_{G\cup H}^{\lceil h/2\rceil} (u,w) + d_{G\cup H}^{\lfloor h/2\rfloor} (v,w)).
$$
Different optimizations of this technique are possible. 

In this paper, we focus only on the time complexity of the case of $h=3$, where we perform the following optimization of query execution. We represent set $H$ as the union of
two (not necessarily disjoint) sets of shortcuts, $H = H_1 \cup H_2$, where an edge belongs to $H_1$ if it is used as the first or third (last) hop on a shortest path in $G\cup H$, and it belongs to $G\cup H_2$ if it is used as the second hop on such a path. By convention, we assume that self-loops at nodes are added to $H_1$, thus e.g. a $3$-hop path between a pair of adjacent nodes in $G$ is constructed by taking a self-loop from $H_1$, the correct edge from $G \subseteq G\cup H_2$, and another self-loop from $H_1$. (Note that we never directly use edges of $G$ as first or last hops in the hopset; if such an edge is required for correctness of construction, it should be explicitly added to set $H_1$.) We further apply an orientation to the shortcuts in $H_1$, constructing a corresponding set of arcs $\vec H_1$, such that, for any node pair $u,v\in V$, there exist $x, y \in V$ such that $(u,x) \in \vec H_1$, $\{x,y\} \in H_2$, $(v,y) \in \vec H_1$, and:
$$
d_G(u,v) = d_G(u,x) + d_G(x,y) + d_G(y,v).
$$
The orientation $(w,z)$ of an arc in $\vec H_1$ indicates that edge $\{w,z\}$ can be used as the first edge of a 3-hop path from $w$ or as the third edge of a 3-hop path to $w$.
We note that $|H_1|\leq |\vec H_1|\leq 2|H_1|$, since each shortcut from $H_1$ corresponds to at most a pair of symmetric arcs in $\vec H_1$. For a node $w\in V$, let $N_1(w) = |\{x \in V : (w,x) \in \vec H_1\}|$ represent the out-neighborhood  of $w$ in the graph $(V,\vec H_1)$. 
To perform shortest path queries on $G$, for each node $w$, we now store the list $\{(x, d_G(w,x)) : x \in N_1(w)\}$. We also store a hash map, mapping all node pairs $\{x,y\} \in H_2$ to the length of the respective link, $d_G(x,y)$. Now, we answer the distance query for a node pair $u,v\in G$ as follows:
$$
d_G(u,v) = \min_{x\in N_1(u), y \in N_1(v) : \{x,y\} \in H_2} (d_G(u,x) + d_G(x,y) + d_G(y,v)).
$$
Using the given data structures, the query is then processed using $|N_1(u)|\cdot |N_1(v)|$ hashmap look-ups, one for each pair $(x,y) \in N_1(u) \times N_1(v)$, i.e., in time $\T_{uv} = O(|N_1(u)|\cdot |N_1(v)|)$. Time $\T_{uv}$ is simply referred to as the \emph{query time} for the considered node pair in the $3$-hopset oracle $H$. Assuming uniform query density over all node pairs, the \emph{uniform-average query time} $\T(H)$ is given as:
$
\T(H) \equiv \E_{uv} \T_{uv} = O\left(\frac1{n^2}\left(\sum_{u \in V}{|N_1(u)|}\right)^2\right) = O(|H_1|^2/n^2).
$
Thus, in the uniform density setting (which we refer to only in Section~\ref{se:lp}), the average time of processing a query is proportional to the square of the average degree of a node with respect to edge set $H_1$. 

The size of set $H_2$ affects only the size of the data structure required by the distance oracle, which is given as at most $\S = O(|E| + |H_1| + |H_2|)$ edges, with each edge represented using $O(\log n)$ bits.

In the 3-hopset distance oracles described in the following sections, we will confine ourselves to describing shortcut sets $H_1$ and $H_2$, noting that the correct orientation $\vec H_1$ of $H_1$ will follow naturally from the details of the provided constructions.

\section{Bounded Skeleton Dimension}\label{se:skeleton}
A formal definition of the notion of skeleton dimension relies on the concept of the geometric realization of a graph, cf.~\cite{KV-SODA2017}. The \emph{geometric realization} $\G$  of $G$ can be seen as the ``continuous'' graph where each edge
is seen as infinitely many vertices of degree two with infinitely small edges, such that for any $uv\in E(G)$ and $t\in [0,1]$, there
is a node in $\G$ at distance $td_G(u,v)$ from $u$ on edge $uv$.
Given a shortest-path tree $T_u$ of node $u$ with length function
$\ell:E(T_u)\rightarrow{\mathbb R^+}$, obtained as the
union of shortest paths $\bigcup \{P_{uv} : v\in V(G)\}$, we treat it as directed
from root to leaves and consider the geometric realization $\Tree_u$ of this directed graph.
We define the \emph{reach} of $v\in V(\Tree_u)$ as the distance from $v$ to the furthest leaf in its subtree of the directed tree $\Tree_u$, i.e., 
$\reach_{\Tree_u}(v) := \max_{x : v \in P_{ux}}d_{\Tree_u}(v,x)$.
For a given value $\alpha>0$, we then define the \emph{skeleton} $T^*_u$ of $T_u$ as the subtree
of $\Tree_u$ induced by nodes with reach at least $\alpha$ times their distance from the root. More precisely, $\Tree^*_u$ is the subtree of $\Tree_u$ induced by
$\{v\in V(\Tree_u)\mid \reach_{\Tree_u}(v) \geq \alpha d_{\Tree_u}(u,v)\}$.

The \emph{$\alpha$-skeleton dimension} $\sk_\alpha$ of a graph $G$ is now
defined as the maximum width of the
skeleton of a shortest path tree, taken over cuts at all possible distances from the root of the tree: $\sk=\max_{u\in V(G)} \max_{r>0}|\cut_r(\Tree^*_u)|$, where $\cut_r(\Tree^*_u)$ is the set of nodes $v\in V(\Tree^*_u)$ with $d_{\Tree^*_u}(u,v) = r$. 
When $\alpha=\frac{1}{2}$, $\sk_{1/2}$ is simply called the \emph{skeleton dimension} of $G$ and we let $\sk=\sk_{1/2}$ denote it.

The definition was originally proposed with $\alpha=\frac{1}{2}$ (for comparison with highway dimension) in the context of USP graphs~\cite{KV-SODA2017}. In the long version~\cite{KV-long2016}, the definition is extended to other choices of $\alpha$ with $0<\alpha<1$ and applies to any choice of shortest paths trees that pairwise agree on their paths (the path from $u$ to $v$ in $T_u$ must be the reverse of the path from $v$ to $u$ in $T_v$). In the non-USP case, the skeleton dimension should be measured with the best choice of agreeing trees.
In particular, if a small perturbation of the edge weights of $G$ provides unique shortest path trees whose skeletons have width at most $\sk_\alpha$, then the skeleton dimension of $G$ is at most $\sk_\alpha$. The $\alpha$-skeleton dimension (with parameter $\alpha$) was introduced in~\cite{KV-long2016} for the sake of a general definition with fixed $\alpha$ value in mind. We use it here in a novel manner with $\alpha$ tending towards 0 as we consider larger distances, enabling analysis of our new construction.

For the definition of the related concept of \emph{highway dimension}, we refer the readers to~\cite{MS-TR}. We note that if a graph $G$ has highway dimension
$h$, then $G$ has skeleton dimension $k=k_{1/2}\leq h$; hence, in all subsequent asymptotic analyses, upper bounds expressed in terms of skeleton dimension can be replaced by analogous bounds in terms of highway dimension.

\subsection{Construction of the 3-Hopset} 

We denote by $L_{\max}$ the maximum length of an edge in graph $G$. The construction of the 3-hopset $H$ is obtained by taking a union of sets of shortcuts, each of which covers sets of node pairs within a given distance range. The first shortcut set $H'$ covers all node pairs $u, v\in V$ with $d_G(u,v) \leq D'$, for some choice of distance bound $D'$, whereas each of the subsequent shortcut sets $H^{(D)}$ covers nodes at a distance in an exponentially increasing distance range, $d_G(u,v) \in [D, D^{1+\eps}]$, where $\eps := \frac{1}{2 \log_2 k}$ is suitably chosen. We then put:
$$
H = H' \cup \bigcup_{i = 1, 2, \ldots} H^{(D'^{\ i(1+\eps)})}.
$$

\subparagraph{Construction of set $H'$. }
We note that a construction of $2$-hopsets for graphs of skeleton dimension $k$ was performed in~\cite{KV-SODA2017}. As a direct corollary of~\cite{KV-SODA2017}[Lem.~2,\ Cor.~1,2], given a distance bound $D'$, there exists a randomized polynomial-time construction of a set of shortcuts $H'$ for graph $G$ with the property that for any pair of nodes $u, v\in V$ with $d_G(u,v) \leq D'$, we have $d^2_{G\cup H'} = d_G(u,v)$, such that $|H'| = O(n k \log D')$, and moreover for all $u \in V$, we have $\E \deg_{H'}(u) = O(k \log D')$ and $\deg_{H'}(u) = O(k \log D' \log \log n + \log n)$. We directly use set $H'$ for the value $D' := L^4_{\max} k^6 \log^{12} n$, considering $H'$ as a $3$-hopset for node pairs $u,v\in V$ with $d_G(u,v) \leq D'$. So we have:
$$|H'| = O(n k (\log \log n + \log L_{\max} + \log k) ),$$ and for all $u \in V$: 
\begin{align*}
\E \deg_{H'}(u) &= O(k (\log \log n + \log L_{\max}  + \log k)),\\
\deg_{H'}(u) &= O(k \log \log n (\log \log n + \log L_{\max}  + \log k) + \log n).
\end{align*}

We remark that, without loss of generality, in asymptotic analysis one may assume that $L_{\max} \leq kL$, where $L$ is the \emph{average} edge length in $G$, noting that edges longer than $kL$ can be subdivided into edges of length at most $kL$ by inserting additional vertices, increasing the number of nodes of the graph only by a multiplicative constant. Thus, in the above bounds, we can replace $(\log \log n + \log L_{\max} + \log k)$ by $(\log \log n + \log L + \log k)$.

\paragraph{Construction of set $H^{(D)}$.} We now proceed to construct a $3$-hopset for node pairs $u,v$ with $d_G(u,v) \in [D, D^{1+\eps}]$. The construction of set $H^{(D)}$ is randomized and completely determined by assignment of real values $\rho(u) \in [0,1]$ to each node $u\in V$, uniformly and independently at random. We condition all subsequent considerations on the event that all values $\rho$ are distinct, i.e., $|\rho(V)| = |V|$, which holds with probability $1$. \mbox{( $\rho(V) = \{\rho(v) | v \in V\}$ )}

Now, hopset $H^{(D)}$ is defined as $H^{(D)} := H^{(D)}_1 \cup H^{(D)}_2$, where following our usual notation, $H^{(D)}_1$ is the set of first and last hops, and $H^{(D)}_2$ is the set of middle hops.

\paragraph{Set of first and last hops.} For $u\in V$, let $R^{(D)}(u)$ be the set of nodes which lie on a shortest path of length at least $D$ which has one of its endpoints at $u$, and which have minimum value of $\rho$ among all vertices on this path at distance in $[D/4,D/2]$ from $u$:
$$
R^{(D)}(u) = \bigcup_{v\in V : d_G(u,v) \geq D} \left\{\argmin_{r \in P_{uv}, d_G(u,r) \in [D/4,D/2]} \rho(r)\right\}.
$$
We now put: $ H^{(D)}_1 := \{ur : u \in V, r \in R(u)\}$.

\paragraph{Set of middle hops.} We put in $ H^{(D)}_2$ links between all pairs of nodes which have a small value of $\rho$, satisfy the natural upper bound of $D^{1+\eps}$ on distance between them, and have sufficiently large reach, i.e., the shortest path between them can be extended by at least $D/4$:
{\small $$
H^{(D)}_2 := \left\{qr : q, r \in \bigcup_{u \in V} R^{(D)}(u) \wedge d_G(q,r) \leq D^{1+\eps} - D/2\wedge (\exists_{v \in V}\ r \in P_{qv} \wedge d_G(r,v) \geq D/4)\right\}.
$$}

The validity of $H$ as a $3$-hopset is immediate to verify from the construction: consider $u,v$ and $i\geq 0$ such that $d_G(u,v) \in [D,D^{1+\eps}]$ with $D=D'^{i(1+\eps)}$. 

For $q=\argmin_{w \in P_{uv}, d_G(u,w) \in [D/4,D/2]} \rho(w)$ and $r=\argmin_{w \in P_{uv}, d_G(u,w) \in [D/2,3D/4]} \rho(w)$, we then have $uq\in H_1^{(D)}$, $qr\in H_2^{(D)}$ and $vr\in H_1^{(D)}$, yielding a 3-hop shortest path from $u$ to $v$. For $d_G(u,v)\le D'$, $H'$ contains a 2-hop shortest path from $u$ to $v$.

\subsection{Bound on 3-Hopset Size and Oracle Time}

\begin{lemma}\label{lem:basic}
Fix $u \in V$ and $D>0$. We have: $|R^{(D)}(u)| \leq k$.
\end{lemma}

\begin{proof}
By the fact that the size of the cut of the skeleton tree for node $u$ at distance $D/2$ from $u$ is upper-bounded by the skeleton dimension $k$, we have that the set of paths $\mathcal{P} := \{\Pi_v : v \in V \wedge  d_G(u,v) \geq D\}$, where $\Pi_v := \{w \in P_{uv} : d_G(u,w) \in [D/4,D/2]\}$, has at least at most $k$ distinct paths, $|\mathcal{P}| \leq k$. The bound on the size of set  
$|R^{(D)}(u)|$ now follows directly from its definition.
\end{proof}

From the above Lemma, it follows that for any $u\in V$, we have $\deg_{H_1^{(D)}}(u) \leq k$. Thus summing over all the $O(\log \log (n L_{\max}) / \log (1 + {\eps})) = O(\log \log (n L_{\max}) \log k)$ levels of the construction, we successively obtain:
{\small
\begin{align}\label{eq:hbound1}
\hspace*{-0.75cm}\deg_{H_1}(u) \leq \deg_{H'}(u) + k \cdot O(\log \log (n L_{\max}) \log k)= O(k \log \log n \log k (\log \log n + \log L) + \log n),\hspace*{-0.5cm}\\
\label{eq:hbound2}
\hspace*{-2.5cm}\E\deg_{H_1}(u) \leq \E\deg_{H'}(u) + k \cdot O(\log \log (n L_{\max}) \log k)= O(k \log k (\log \log n + \log L)),\hspace*{-0.5cm}\\
\label{eq:hbound3}
|H_1| \leq |H'| +  nk \cdot O(\log \log (n L_{\max}) \log k) = O(nk \log k(\log \log n + \log L)).\hspace*{-0.5cm}
\end{align}}

We now proceed to bound the size of the set $H_2$ of middle hopsets.
\begin{lemma}\label{lem:bla1}
Fix $D \geq D'$. With probability $1 - O(1/n^2)$, it holds that for all $u\in V$ and for all $r \in R^{(D)}(u)$, we have $\rho(r) \leq L_{\max} / D$.
\end{lemma}

\begin{proof}
As noted in the proof of Lemma~\ref{lem:basic}, to be included in $R^{(D)}(u)$, a node $r$ must be the minimum element along one of the at most $k$ possible paths $\Pi_v$. Each such path includes all nodes on the path $P_{uv}$ at distance in the range $[D/4, D/2]$ from $u$, where we recall that $D \geq D' = C L^4_{\max} k^6 \log^{12} n > C L_{\max} \ln^2 n$, for some sufficiently large choice of constant $C > 0$. It follows that each path $\Pi_v$ contains $|\Pi_v|\geq \max\{\ln^2 n, \frac{CD}{8L_{\max}}\}$ nodes. Now, taking note of the independence of the choice of random variables $(\rho(w) : w \in \Pi_v)$, we have by a simple concentration bound that $\Pr[\min \rho(\Pi_v) > L_{\max}/D] \leq O(1/n^4)$, for a suitable choice of constant $C$. By taking a union bound over all paths $\Pi_v$ in $\mathcal{P}$, and then another union bound over all $u \in V$, the claim follows.
\end{proof}

We now proceed  under the assumption that the event from the claim of the Lemma holds.  We now consider an arbitrary node $q \in R^{(D)}(u)$ for some $u \in V$, and look at $\deg_{H_2^{(D)}}(q)$. We now have that if $qr \in H_2^{(D)}$, then by the definition of $ H_2^{(D)}$ and the above Lemma, the following conditions jointly hold:
\begin{itemize}
\item $\rho(r) \leq L_{\max} / D$
\item $r \in \{w \in V : \exists_{v \in V}\ D^{1+\eps} \geq d_G(q,v) \geq d_G(q,w) + D/4 \wedge P_{qw} \subseteq P_{gv}\} =: W(q)$.
\end{itemize}
We note that $W(q)$ is the subset of the vertex set of the shortest path tree of node $q$, pruned to contain only those paths which have reach at least $D/4$ at depth less than $D^{1 + \eps}$. This tree has depth bounded by $D^{1 + \eps}$, and width bounded by an $\alpha$-skeleton dimension $k_{\alpha}$ (following~\cite{KV-long2016}), with parameter $\alpha = \frac{D/4}{D^{1+\eps}} = D^{-\eps}/4$. Following~\cite{KV-long2016}[Section 6], $k_{\alpha}$ can be easily expressed using skeleton dimension $k=k_{1/2}$ as:
$$
k_{\alpha}\le k^{\lceil{\log_2(1+1/\alpha)}\rceil} < k^{1 + \log_2 (4 D^{\eps})} = k^3 D^{\eps \log_2 k}.
$$
We then have $|W(q)| \leq D^{1 + \eps} k_{\alpha} < k^3 D^{1 + \eps (1+\log_2 k)}$. Moreover, by an easy concentration bound, we have that for all $q \in V$, $|\{r \in W(q) : \rho(r) \leq L_{\max} / D\}|  = O(\log n) + \frac{2 L_{\max}}{D}|W(q)|$, with probability $1 - O(1/n^2)$.  It follows that with probability $1-O(1/n^2)$, we have for all $q \in \bigcup_{u \in V}R^{(D)}(u)$:
$$
\deg_{H_2^{(D)}}(q) \leq O(\log n) + \frac{2 L_{\max}}{D}|W(q)| \leq  O(\log n + L_{\max} k^3 D^{\eps \log_2 k}).
$$
Noting that with probability $1 - O(1/n^2)$:
$$
|\bigcup_{u \in V}R^{(D)}(u)| \leq |\{w \in V : \rho(w) \leq L_{\max}/D| \leq O(\log n + n L_{\max}/D)
$$
we finally obtain that with probability $1 - O(1/n^2)$:
{\small
\begin{align*}
|H_2^{(D)}| &\leq O(\log n + n L_{\max}/D) O(\log n + L_{\max} k^3 D^{\eps \log_2 k}) = O(\log^2 n +  nL^2_{\max} k^3 D^{\eps \log_2 k - 1})\\
& \leq O(nL^2_{\max} k^3 D^{-1/2}) \leq O(n D'^{-1/4})  \leq O(n/\log^3 n),
\end{align*}}
where in the last two transformations we use the fact that $\eps = \frac{1}{2\log_2 k}$ and that $D \geq D' \geq L^4_{\max} k^6 \log^{12} n$.
Using a union bound and summing over all levels of the construction, we eventually obtain that with probability $1 - O(1/n)$:
\begin{equation}\label{eq:hbound4}
|H_2| \leq O(n /\log_2 n).
\end{equation}
Thus, the set of middle links is sparse and does not contribute to the asymptotic size of the overall representation of the $3$-hopset.

Overall, considering a randomized construction which rejects random choices of $\rho$ for which any of the considered w.h.p.\ events fail, by combining Eq.~\eqref{eq:hbound1}--\eqref{eq:hbound4} with the hopset-based distance oracle framework described in the Preliminaries, we obtain the following Theorem.

\begin{theorem}
For a unique shortest path graph with skeleton dimension $k$ and average link length $L\geq 1$, there exists a randomized construction of a $3$-hopset distance oracle of size $|H| = O(nk \log k (\log \log n + \log L))$, which for an arbitrary queried node pair performs distance queries in expected time $O(k^2 \log^2 k(\log^2 \log n + \log^2 L))$ (where the expectation is taken over the randomized construction of the oracle), and in time $O(k^2  \log^2 k \log^2 \log n(\log^2 \log n + \log^2 L) + \log^2 n)$ with certainty.
\end{theorem}

In particular, for graphs with constant-length edges and small skeleton dimension ($k = O(\log n)$), the $3$-hopset has size  $|H| = O(nk \log k\log \log n)$, with expected time of any query given as $O(k^2 \log^2 k\log^2 \log n)$.

\section{LP-based Approximation Algorithm}\label{se:lp}

In this section, we propose an Integer Linear Programming (ILP) formulation for $h$-hopsets with a minimum number of edges, which we then relax to a LP formulation. Whereas both formulations are applicable to the general case, we  prove relations between them only for USP graphs.

\subsection{ILP and LP Formulations}\label{se:ilp}
A necessary and sufficient condition for $H$ to be a $h$-hopset for $G$ is that for every pair of vertices $s, t$ there exists a path $P_{st} = (s = v_0, v_1, \ldots, v_{l_{st}} = t)$ in $G \cup H$ such that $l_{st} \leq h$ and in graph $G$ there exists some shortest $s-t$ path passing through all of the vertices $v_0, \ldots, v_{l_{st}}$, in the given order. For a fixed pair $s, t$, we consider the directed graph $H^{st}$ with vertex set $V \times \{0, \ldots, h\} \equiv V_h$ (by convention, elements of $V_h$ will be denoted compactly as $v_i$, where $v\in V$, $i\in \{0, \ldots, h\}$) and with an arc set defined as follows. For $i \in \{0,\ldots, h-1\}$, we add arc $(u_i, v_{i+1})$ to $H^{st}$ if and only if $\{u, v\} \in G\cup H$ and $u,v$ lie on some shortest $s-t$ path in the given order, i.e., if $d_G(s,u) + d_G(u,v) + d_G(v,t) = d_G(s,t)$. In particular, all arcs of the form $(u_{i}, u_{i+1})$, for $u\in V$ on a $s-t$ shortest path, belong to $H^{st}$. Now, we have that $H$ is a $h$-hopset for $G$ if and only if there exists a path from $s_0$ to $t_h$ in $H^{st}$. This is equivalent to saying that for all $s,t \in V$, the flow value from $s_0$ to $t_h$ is at least $1$ in $H^{st}$. Given graph $G$, we thus have the following ILP formulation for the minimum $h$-hopset problem, using indicator variables $x_{uv}$ for $G \cup H$ (given as $1$ if $\{u,v\} \in G \cup H$ and $0$ otherwise) and variables $f^{st}_{u_iv_j}$, representing the flow value along arc $(u_i,v_j)$ in $H^{st}$:
\begin{align}
\text{Minimize:\quad\quad\quad\quad\quad\quad} &
\sum_{u \neq v, \{u,v\} \notin E} x_{uv}\label{eq:lp1start}\\
\intertext{\quad \quad Subject to:}
x_{uv} &\in \{0,1\}\\
0 \leq f^{st}_{u_iv_j} &\leq \begin{cases}
x_{uv}, & \text{\small if $j = i+1$ and $d_G(s,u) + d_G(u,v) + d_G(v,t) = d_G(s,t)$,}\\
0, &\text{otherwise}.
\end{cases}\\
\sum_{u_i} f^{st}_{v_ju_i} - \sum_{u_i} f^{st}_{u_iv_j}  &= \begin{cases}0,& \text{for $v_j \in V_h \setminus \{s_0, t_h\}$}\\
+1,& \text{for $v_j = s_0$}\\
-1,& \text{for $v_j = t_h$}
\end{cases},\label{eq:lp1end}\end{align}
where indices $s, t, u, v$ traverse $V$ and indices $i, j$ traverse $\{0,\ldots, h\}$.

To obtain an LP relaxation of the above problem, we replace the integral condition $x_{uv}\in \{0,1\}$ by the fractional one $x_{uv}\in [0,1]$. We look at the connection between the integral and fractional forms for the special case of unique shortest path graphs.

We remark that the above formulation can be seen as a generalization of the LP and ILP statement of Angelidakis et al.~\cite{AngelidakisMO17} 
proposed for the special case of 2-hop labeling. In the case of 2-hop labeling, Angelidakis et al. do not rely on an explicit flow formulation but use a single constraint of the simpler form $\sum_{w \in P^{st}} \min\{x_{sw}, x_{wt}\} \geq 1$, where $P^{st}$ represents the set of nodes on some shortest $s-t$ path in $G$. However, the analysis of the integrality gap does not carry over from the case of $h=2$ to $h>2$, i.e., as soon as there exist internal shortcuts which have neither $s$ nor $t$ as one of their endpoints.

\subsection{Bounding Integrality Gap for Unique Shortest Path Graphs}\label{se:igap}

We analyze the integrality gap of the above LP formulation for the case of \emph{unique shortest path (USP) graphs}, i.e., graphs in which each pair of nodes $s,t \in V$ is connected by a unique shortest path $P^{st}$ in $G$. We will occasionally identify $P^{st}$ with its set of nodes, and we will introduce a linear order on its vertices, writing for $u, v \in P^{st}$ that $u <^{st} v$ if $d_G(s,u) < d_G(s,v)$; we will denote the order simply as ``$<$'' when the path $P^{st}$ is clear from the context. Observe that in the LP formulation, we may have $f^{st}_{u_iv_j} \neq 0$ only if $u <^{st} v$ and $j = i+1$. Thus, fixing $s, t\in V$, the flow $f^{st} =(f^{st}_{u_iv_j} : u_i, v_j \in V_h)$ is non-zero between vertices of $\{P^{st}\}\times\{0,1,\ldots,h\}$ only, and the flow is oriented towards $t$ on this path.

Let $(x_{uv},f^{st}_{u_iv_j})$ be a fixed solution to the LP problem in a USP graph, with cost $\costlp = \sum_{u \neq v, \{u,v\} \notin E} x_{uv}$. We will show how to use this set to construct a valid hopset $H''$ for $G$  (thus, equivalently, also solving the ILP formulation).
We first apply a randomized rounding procedure following the classical scheme of Raghavan and Thomson~\cite{RaghavanT87}. We define the family of independent random variables $(x'_{u_iv_{i+1}} : u,v\in V, i\in\{0,\ldots,h\})$, with $x'_{u_iv_{i+1}} \in \{0,1\}$. For $u \neq v, \{u,v\} \notin E$ we put $\Pr[x'_{u_iv_{i+1}} = 1] = \min\{C x_{uv}, 1\}$, where $C \geq 1$ is a suitably chosen probability amplification parameter (we put $C = 8h \ln n$). We will assume, without affecting the validity or cost of the solution, that $x_{uv} = x'_{u_iv_{i+1}} = 1$, when $u=v$ or $\{u,v\} \in E$.

We denote $H' = \{\{u,v\}: u,v \in V \wedge u\neq v \wedge \{u,v\} \notin E \wedge \exists_{i \in \{0,\ldots,h-1\}}\ x'_{u_iv_{i+1}} = 1\}$. Let $\pi : V \to \{1,\ldots,n\}$ be a bijection picked uniformly at random (it is a random permutation when $V=\{1,\ldots,n\}$). We define the set of shortcuts $S(\{u,v\})$ associated with each pair $\{u,v\} \in H'$ as the set of all pairs of nodes on path $P^{uv}$, one of which is a prefix minimum on this path with respect to $\pi$, and the other of which is a suffix minimum with respect to $\pi$:
{\small $$
S(\{u,v\}) := \left\{\{u^*,v^*\}: u^*, v^* \in P^{uv} \wedge \pi(u^*) = \min_{z \in P^{uv}, z \leq^{uv} u^*} \pi(z) \wedge \pi(v^*) = \min_{z \in P^{uv}, z \geq^{uv} v^*} \pi(z)\right\}.
$$}
The obtained solution is given as the set of all such shortcuts:
$
H'' := \bigcup_{\{u,v\} \in H'} S(\{u,v\}).
$
\begin{proposition}\label{thm:uspgap}
With probability $1 - O(1/n)$, set $H''$ is a hopset for $G$ of size $O(h^2\log^3 n \cdot \costlp)$.
\end{proposition}

The rest of the section is devoted to the proof of Proposition~\ref{thm:uspgap}. 

\subsubsection{Size of the Hopset $H''$}

\begin{proposition}\label{prop:Hbis}
We have $|H''| = O(h^2\log^3 n \cdot \costlp)$, with probability $1 - O(1/n)$.
\end{proposition}
\begin{proof}
We first remark that for any $i\in\{0,\ldots,h-1\}$, by a standard application of a multiplicative Chernoff bound, we have that the following bound holds with probability $1 - O(1/n^2)$:
$$
\sum_{u\neq v, \{u,v\}\notin E} x'_{u_iv_{i+1}} \leq 2C \sum_{u\neq v, \{u,v\}\notin E} x_{u_iv_{i+1}} = O(h \log n \cdot \costlp)
$$
It follows by a union bound over $i\in \{0,\ldots,h-1\}$ that $|H'| = O(h^2 \log n \cdot \costlp)$, with probability $1 - O(1/n)$.

We now proceed to bound the size of each set $S(\{u,v\})$, for $\{u,v\} \in H'$. This is given precisely by the product of the size of the set of prefix minima and suffix minima of permutation $\pi$ on path $P^{uv}$. Denoting the random variable describing the number of prefix minima on a path as $X^{st} := |\{u^* \in P^{st} : \pi(u^*) = \min_{z \in P^{st}, z \leq^{st} u^*}\}|$, we have:
$$
|S(\{u,v\})| = X^{uv} \cdot X^{vu}.
$$
It is well-known the number of prefix minima has expectation $\E X^{uv} = \ln |P^{uv}| + O(1) \leq \ln n + O(1)$ and that the distribution of $X^{uv}$ is concentrated around its expectation; in particular, by a simple multiplicative Chernoff bound, we have that $\Pr[ X^{uv} \leq 4 \ln n] \geq 1 - n^{-3}$. Applying a union bound over all $\{u,v\}$, we have:
$$
\Pr[\forall_{\{u,v\}\in H'}\ |S(\{u,v\})| \leq 16 \ln^2 n] \geq 1 - n^{-1}.
$$
Overall, we thus have that $|H''| = O(h^2 \log n \cdot \costlp \cdot \log^2n )  = O(h^2 \log^3 n \cdot \costlp)$, with probability $1 - O(1/n)$.
\end{proof}

\subsubsection{Correctness of the Hopset $H''$}

For fixed $s,t \in V$, the choice of $x'_{u_iv_{i+1}}$ is performed iteratively over $i$, as a random process. Each step $i = 0,1,\ldots, h-1$ of this process determines the vertex $v^{(i+1)st} \in P^{st}$, given inductively as:
$$
v^{(i+1)st} = \max_{(<^{st})}
\{v \in P^{st} : \exists_{u \in P^{st}}\  u \leq v^{(i)st} \wedge x'_{u_iv_{i+1}} =1  \},
$$
where we denote $v^{(0)st} := s$.

First of all, observe that we have the following sufficient condition for the validity of a $h$-hopset for the pair $s,t$.
\begin{lemma}\label{lem:cond}
If $v^{(h)st} = t$, then there exists a $s-t$ path in $G \cup H''$ with at most $h$ hops whose vertices form an increasing subsequence on $P^{st}$ according to the order ``$<^{st}$''.
\end{lemma}
\begin{proof}
For $i \in \{0,\ldots,h-1\}$, denote by $u^{(i+1)st}$ the vertex $u$ used in the definition of $v^{(i+1)st}$, i.e.:
$$u^{(i+1)st} = \max_{(<^{st})} \{u \in P^{st} : x'_{u_i v^{(i+1)st}_{i+1}} =1 \}.$$
Note that $u^{(i+1)st} \leq v^{(i)st} \leq v^{(i+1)st}$.
For some $l\leq h$, let $(\phi_0, \ldots, \phi_l) \subseteq (0,\ldots,h)$, with $\phi_0 = 0$ and $\phi_l = h$, denote a minimal subsequence of indices such that $u^{(\phi_i)st} \leq v^{(\phi_{i-1})st} \leq u^{(\phi_{i+1})st} \leq v^{(\phi_i)st}$, for all $i \in \{1,\ldots,l-1\}$.
Note that each path $P^{u^{(\phi_{i+1})st} v^{(\phi_i)st}}$ is a subpath of $P^{st}$ by the unique shortest path condition, and consider the minimum vertex $z^{(i)st}$ according to permutation $\pi$ on this subpath: $z^{(i)st} := \arg\min \{\pi(z) : z \in P^{st} \wedge u^{(\phi_{i+1})st} \leq z \leq v^{(\phi_i)st} \}$. Note that $z^{(0)st}=s$, $z^{(l)st}=t$, and for all $i \in \{0,\ldots, l-1\}$, we have:
$$
u^{(\phi_{i+1})st} \leq z^{(i)st} \leq v^{(\phi_i)st} \leq u^{(\phi_{i+2})st} \leq z^{(i+1)st} \leq v^{(\phi_{i+1})st}.
$$
We have $\{u^{(\phi_{i+1})st},v^{(\phi_{i+1})st}\} \in H'$, and moreover $z^{(i)st}$ is a prefix minimum with respect to $\pi$ on $P^{u^{(\phi_{i+1})st}v^{(\phi_{i+1})st}}$ (for the subpath $P^{u^{(\phi_{i+1})st}v^{(\phi_i)st}}$), whereas $z^{(i+1)st}$ is a prefix maximum with respect to $\pi$ on $P^{u^{(\phi_{i+1})st}v^{(\phi_{i+1})st}}$ (for the subpath $P^{u^{(\phi_{i+2})st}v^{(\phi_{i+1})st}}$). It follows from the definition of $H''$ that $\{z^{(i)st},z^{(i+1)st}\} \in H''$. Recalling that $z^{(i)st} \leq^{st} z^{(i+1)st}$, $z^{(0)st} = s$ and $z^{(l)st} = t$ for some $l\leq h$, the claim follows from the existence of the path $(z^{(0)st}, z^{(1)st}, \ldots, z^{(l)st})$.
\end{proof}
The rest of the proof of correctness is devoted to showing that the event ``$v^{(h)st} = t$'' holds with high probability. We have the following claim.
\begin{lemma}\label{lem:xconc}
$$\Pr\left[\sum_{u, v \in P^{st}:\ u\leq v^{(i)st},\ v > v^{(i+1)st}} x_{uv} > \frac{1}{2h}\right] < n^{-4}.
$$
\end{lemma}
\begin{proof}
Denote the probability from the claim by $p$.
Conditioned on the choice of $v^{(1)st}, \ldots,v^{(i)st}$, at the beginning of step $i$, let $w$ be the right-most (largest) vertex on path $P^{st}$ such that
$$
\sum_{u, v \in P^{st}:\ u\leq v^{(i)st},\ v > w} x_{uv} > \frac{1}{2h}.
$$
Directly by the definition of $v^{(i+1)st}$, we have:
\begin{align*}
p &= \Pr[v^{(i+1)st} \leq w] = \Pr \left[\forall_{u, v \in P^{st}:\ u\leq v^{(i)st},\ v > w} \ x'_{u_iv_{i+1}} = 0 \right] = \prod_{u, v \in P^{st}:\ u\leq v^{(i)st},\ v > w} \!\!\!\!\! \max\{0,1-C x_{uv}\} \\ & \leq \prod_{u, v \in P^{st}:\ u\leq v^{(i)st},\ v > w} \!\!\!\!\! (1-C x_{uv}) \leq \exp\left[-\sum_{u, v \in P^{st}:\ u\leq v^{(i)st},\ v > w} Cx_{uv}\right] < e^{-C/2h} = n^{-4}.
\end{align*}
\end{proof}

We now consider the graph $H^{st}$ inferred from the (not necessarily integral) solution to the LP, given on vertex set $V$ as the set of edges $uv$, such that $f^{st}_{u_jv_{j+1}} > 0$ for some $j$.

Each step $i = 0,1,\ldots, h-1$ of the considered process of random choice determines the following $s_0-t_h$-flow $F^{(i+1)st}$ on an edge-weighted version of graph $H^{st}$, described by its flow value $f^{(i+1)st}_{u_j v_{j+1}}$ on each arc $(u_j, v_{j+1})$ of $H^{st}$ as follows. 
$F^{(i+1)st}$ is set as a maximum $s_0-t_h$ flow (with ties broken deterministically in an arbitrary manner) in an edge-weighting of $H^{st}$ such that the capacity of arc $(u_j, v_{j+1})$ is $f^{(i)st}_{u_j v_{j+1}}$, for all arcs of $H^{st}$, except for arcs $(u_i, v_{i+1})$ with $v > v^{(i+1)st}$, whose capacity is set to $0$. By convention, we denote $f^{(0)st}_{u_j v_{j+1}} := f^{st}_{u_j v_{j+1}}$, i.e., as the flow value on the considered arc in the optimal solution to the LP.

Denote by $|F^{(i)st}|$ the value of flow $F^{(i)st}$. The following claim holds.
\begin{lemma}\label{lem:fconc}
$\Pr[|F^{(i+1)st}| \geq |F^{(i)st}| - \frac{1}{2h}] \geq 1 - n^{-4}.$
\end{lemma}
\begin{proof}
We note that for any $u,v \in P^{st}$ such that $u  > v^{(i)st}$ we have $f^{(i+1)st}_{u_i v_{i+1}} = 0$, since in the $i$-th step of the considered process, the in-capacity of vertex $u_i$ is given as $\sum_{w \in P^{st}} f^{(i)st}_{w_{i-1}u_i} = 0$ by the definition of the $(i-1)$-st step of the process.

Moreover, for any arc $(u_j, v_{j+1})$ of $H^{st}$, the values $f^{(i)st}_{u_j v_{j+1}}$ are clearly non-increasing with $i$, thus in particular:
$$
f^{(i)st}_{u_j v_{j+1}} \leq f^{(i-1)st}_{u_j v_{j+1}}  \leq \ldots \leq f^{(0)st}_{u_j v_{j+1}} \leq x_{uv}.
$$

Combining the two above observations, by comparing the size of any two cuts in graph $H^{st}$ for its weightings in successive steps and taking into account the above observations, we obtain the following expression which is used to lower-bound $|F^{(i+1)st}|$:
\begin{align*}
|F^{(i)st}| - |F^{(i+1)st}|  &\leq\sum_{u, v \in P^{st}:\ v > v^{(i+1)st}} f^{(i)st}_{u_i v_{i+1}} = \sum_{u, v \in P^{st}:\ u\leq v^{(i)st},\ v > v^{(i+1)st}}  f^{(i)st}_{u_i v_{i+1}} \\
&\leq \sum_{u, v \in P^{st}:\ u\leq v^{(i)st},\ v > v^{(i+1)st}} x_{uv}
\end{align*}

Thus, applying Lemma~\ref{lem:xconc} we obtain the claim.
\end{proof}
\begin{lemma}
$\Pr[v^{(h)st} = t] \geq 1 - n^{-3}$.
\end{lemma}
\begin{proof}
First note that if $v^{(h)st} \neq t$, then $v^{(h)st} < t$, and it follows that $|F^{(h)st}|=0$ because all the capacities of arcs entering node $t_h$ are equal to $0$ by definition in the graph in which flow $F^{(h)st}$ is considered.

Now, observe that using Lemma~\ref{lem:fconc} and  applying a union bound over $i$, we obtain:
$\Pr[|F^{(h)st}| \geq |F^{(0)st}| - \frac{h}{2h}] \geq 1 - hn^{-4} \geq 1 - n^{-3}$.
Observe next that $F^{(0)st} \geq 1$ by the constraints of the LP solution, hence $|F^{(h)st}|$ is strictly positive with probability at least $1 - n^{-3}$.
\end{proof}
Applying a union bound over all pairs $s,t\in V$, we obtain $\Pr[\forall_{s,t\in V} v^{(h)st} = t] \geq 1 - n^{-1}$. The correctness of the scheme with probability $1 - n^{-1}$ follows directly from Lemma~\ref{lem:cond}.
 
We remark that the Proposition~\ref{thm:uspgap} implies that the $h$-hopset problem can be efficiently approximated by finding an optimal fractional LP solution and constructing set $H''$.

\begin{theorem}
There exists a randomized polynomial-time $O(\poly\log n)$-approximation algorithm for the $h$-hopset problem in unique shortest path graphs, for any $h \leq O(\poly\log n)$.
\end{theorem}

\enlargethispage{3.5mm}\vspace{-3.5mm}
\subsection{Approximating Average Query Time for 3-Hopsets}\label{se:lp3}

In order to design an efficient distance oracle based on $3$-hopsets, we follow the framework described in the preliminaries and use an LP-rounding technique to obtain sets $H_1 \cup H_2 =: H$.
The obtained claim relies on the notion of uniform-average query time introduced in the Preliminaries. 

\begin{theorem}\label{th:3tradeoff}
For any feasible bound $\S$, let $H_{OPT,\S}$ be a $3$-hopset for a unique shortest path graph, which satisfies the given bound on the number of edges $|H_{OPT,\S}| \leq \S$ and such that the uniform-average query time $\T(H_{OPT,\S})$ is minimized.
Then, there exists a randomized polynomial-time algorithm which finds a $3$-hopset $H$ with  $|H''| \leq O(\log^3 n)\S$ and $\T(H'') \leq O( \log^4 n) \T(H_{OPT,\S})$.
\end{theorem}

\begin{proof}
In this case, for the ILP statement we associate with each edge $uv$ a binary indicator variable $x_{uv}^{(1)} \in \{0,1\}$ stating if $uv \in H_1$, and a second indicator variable $x_{uv} \in \{0,1\}$, with $x_{uv} \geq x_{uv}^{(1)}$, stating if $uv \in H_1 \cup H_2$. The problem of minimizing the query time of the oracle with size bound $\S$ for uniform node-pair query frequencies is now given as (compare with~\eqref{eq:lp1start}--\eqref{eq:lp1end}):

\begin{align}
\intertext{Minimize:}
\sum_{u \neq v, \{u,v\} \notin E} x^{(1)}_{uv}\label{eq:lp2start}
\intertext{Subject to:}
\sum_{u \neq v, \{u,v\} \notin E} x_{uv} &\leq \S\\
0 \leq f^{st}_{u_iv_j} &\leq \begin{cases}
x^{(1)}_{uv}, & \text{if $2\neq j = i+1$ and $d_G(s,u) + d_G(u,v) + d_G(v,t) = d_G(s,t)$,}\\
x_{uv}, & \text{if $2 = j = i+1$ and $d_G(s,u) + d_G(u,v) + d_G(v,t) = d_G(s,t)$,}\\
0, &\text{otherwise}.
\end{cases}\\
\sum_{u_i} f^{st}_{v_ju_i} - \sum_{u_i} f^{st}_{u_iv_j}  &= \begin{cases}0,& \text{for $v_j \in V_h \setminus \{s_0, t_h\}$}\\
+1,& \text{for $v_j = s_0$}\\
-1,& \text{for $v_j = t_h$}
\end{cases},\end{align}
and its LP relaxation on variables $x_{uv}, x^{(1)}_{uv}$ takes the form of the constraint:
$$0 \leq x^{(1)}_{uv} \leq x_{uv} \leq 1,$$
where as usual indices $s, t, u, v$ traverse $V$ and indices $i, j$ traverse $\{0,1,2,3\}$.

The construction of the integral hopset $H''$ based on the LP solution takes place as in the previous Subsection (for the case of $h=3$), with the exception that for the first and last (third) hop, variables $x^{(1)}_{uv}$ should be used in place of $x_{uv}$ in the construction. By an analogue of Proposition~\ref{prop:Hbis}, we have $|H''| = O(\S \log^3 n)$, with high probability. We consider the natural decomposition $H'' := H''_1 \cup H''_2$ according to the number of the used hop along the path, and obtain by a similar (straightforward) concentration analysis that for all $u \in V$:
$$
\deg_{H_1''}(u) \leq O(\log^2 n) \sum_{v \in V} x^{(1)}_{uv}.
$$
and so, computing the sum of degrees over all $u$:
$$
|H_1''| \leq O(\log^2 n) \sum_{v \in V} x^{(1)}_{uv}.
$$
Noting that the sum on the right-hand side is precisely the minimization criterion in the LP formulation~\eqref{eq:lp2start}, we obtain the claim of the theorem.
\end{proof}
We remark that the above Theorem can be directly generalized to a notion of average query time for non-uniform query densities, in which the goal is to minimize expected query time in a model in which each node $v\in V$ is assigned its relative frequency $f_v \in [0,1]$, and a node pair $uv$ is queried with frequency $f_u f_v$.

\section{Bounded Treewidth Graphs}\label{se:boundedtw}
\label{se:tree}\label{se:acker}\label{se:btw}

We now show how to obtain $h$-hopsets for graphs with bounded treewidth by following a classical construction for trees. We first begin with preliminaries recalling the definitions of treewidth and inverse Ackermann function.

\paragraph{Treewidth definition.}
Recall that a graph $G$ has treewidth $t$ if there exists a tree $T$ whose nodes are subsets of $V(G)$ called \emph{bags} such that: $|X|\le t+1$ for all $X\in V(T)$; for all edges $uv\in E(G)$, there exists a bag $X\in V(T)$ containing both $u$ and $v$ ($u,v\in X$); and for all nodes $u\in V(G)$, the bags containing $u$ form a sub-tree of $T$. Without loss of generality, we assume that each bag contains exactly $t+1$ nodes, and that two neighboring bags share exactly $t$ nodes (the decomposition is standard). This implies $|V(T)|\le n$ as each bag brings one new node.  Note that removing a non-leaf bag separates the graph into several connected components. We consider that all edges of $T$  have weight 1. For convenience, we assume that $T$ is rooted at some bag $R$ and define for each node $u\in V(G)$ the root bag of $u$ as the bag $R_u\in V(T)$ containing $u$ which is closest to the root.

\paragraph{Ackermann notation.}
Following \cite{AlonSchieber}, we introduce the following variants of the Ackermann function:\\
\[
\left\{\begin{array}{ll}
  A(0,j)=2j, \mbox{ for } j \ge 0\\
  A(i,0)=1, \mbox{ for } i \ge 1\\
  A(i,j)=A(i-1,A(i,j-1)), \mbox{ for } i,j \geq1;\\
\end{array}\right.
\hfill\mbox{ and }
\left\{\begin{array}{ll}
  B(0,j)=j^2, \mbox{ for } j \ge 0\\
  B(i,0)=2, \mbox{ for } i \ge 1\\
  B(i,j)=B(i-1,B(i,j-1)), \mbox{ for } i,j \geq1.\\
\end{array}\right.
\]
The \emph{$k$th-row inverse Ackermann} function $\lambda_k(.)$ is defined by $\lambda_{2i}(n)=\min\{ j \mid A(i,j)\geq n \}$ and $\lambda_{2i+1}(n)=\min\{ j \mid B(i,j) \geq  n \}$ for $i\ge 0$. Equivalently, we have (up to integer ceiling) $\lambda_0(n)=\frac{n}{2}$, $\lambda_1(n)=\sqrt{n}$ and $\lambda_k(n) = \lambda_{k-2}^*(n)$ where we define for any function $f$: $f^{(0)}(n)=n$, $f^{(i)}(n)=f(f^{(i-1)}(n))$ for $i>0$, and $f^*(n)=\min\{ j \mid f^{(j)}(n)\le 1\}$.
Note that $\lambda_2(n)=\log n$, $\lambda_3(n)=\log \log n$, $\lambda_4(n)=\log^*n$ and $\lambda_5(n)=\frac{1}{2}\log^*n$ (we omit integer ceilings).
The \emph{inverse Ackermann} function is defined as $\alpha(n)=\min\{ j \mid A(j,j)\ge n \}$. Note that we have $\lambda_{2\alpha(n)}(n)=\alpha(n)$.

\medskip

We first consider the case of (weighted) trees for which 
the construction of $h$-hopsets is classical (even though the connection with hopsets was not made). It is implicit in \cite{AlonSchieber,Chazelle87a}, explicit for unweighted trees in~\cite{BodlaenderTS94} and directed trees in~\cite{Thorup97}. We provide a short construction which fine-grains the dependence of the hopset size on $h$ (e.g., replacing $2h$ by $h$ with respect to the asymptotic analysis in \cite{AlonSchieber}). The construction is  based on the following folklore lemma for splitting a tree into smaller sub-trees (it can be seen as a generalization of the existence of a centroid).

\begin{lemma}
\label{lem:splittree}
Given a rooted tree $T$ with $n$ nodes and a value $p>1$, there exists a set $P$ of at most $2p$ nodes such that each connected component of $T\setminus P$ contains less than $n/p$ nodes and is connected to at most two nodes in $P$. Set $P$ can be computed in linear time through a bottom-up traversal of the tree.
\end{lemma}

\begin{proof}
Start with $P'=\emptyset$ and root $T$ at some arbitrary node $r$. As long as the connected component of $T\setminus P'$ containing root $r$ has $n/p$ nodes or more, add to $P'$ a node $u$ from this component such that the subtree $T(u)$ rooted as $u$ has size $n/p$ or more while $|T(v)|<n/p$ for all descendants $v$ of $u$. This results in a set $P'$ having at most $p$  nodes such that the connected components of $T\setminus P'$ have size less than $n/p$. Define $P''$ as the set of lowest common ancestors of any two nodes $u,v\in P'$. The size of $P''$ is at most $p-1$ since its nodes correspond to the internal nodes with two children or more in the minimal sub-tree containing $P'$ which has at most $p$ leaves. Let $P=P'\cup P''$ be the union of $P'$ and $P''$. For any connected component $T'$ of $T\setminus P$, there exist at most two nodes in $P$ that are connected to nodes of $T'$ in $T$: at most one is connected to the root $r'$ of $T'$ ($T'$ is considered as a sub-tree of $T$) and at most one has its parent in $T'$ (if there were two such nodes, their lowest common ancestor would be in $P$ and not in $T'$, contradicting the connectivity of $T'$).
\end{proof}

\paragraph{$h$-hopset construction for trees.}
A 1-hopset in a tree $T$ is obtained by adding all pairs as edges with appropriate weight. For $h > 1$, we recursively define a $h$-hopset of $T$ as follows. Select a set $P$ of $2p$ nodes at most  with $p=\frac{n}{\lambda_{h-2}(n)}$ according to Lemma~\ref{lem:splittree}. 
When $h=2$, we add an edge from each node $u$ of $T$ to each node in $P$. When $h>2$, we consider the forest $T'$ induced by nodes in $P$: it has node set $P$ and edges $xy$ such that $y$ is the closest ancestor of $x$ in $T$ that belongs to $P$. The weight of such an edge is defined as $w'(x,y)=d_T(x,y)$. We then add a $(h-2)$-hopset of $T'$ to the construction. Additionally, we add one or two edges per node not in $P$: for each  connected component $C$ of $T\setminus P$, add an edge $ux$ for each node $u\in C$ and each $x\in P$ connected to $C$. Note that Lemma~\ref{lem:splittree} ensures that there are at most two such nodes $x$ for a given component $C$. In both cases ($h\ge 2$), we construct recursively a $h$-hopset of each sub-tree induced by a connected component $C$ of $T\setminus P$. In the special case of $h=3$, the $(h-2)$-hopsets contribute to $H_2$ while all edges connecting to a node in some selected set $P$ contribute to $H_1$ according to the $H = H_1 \cup H_2$ convention introduced in the Preliminaries.

We now state the parameters of the designed hopset for trees.
\begin{proposition}\label{prop:tree}
For any integer $h>1$ and weighted tree $T$ with $n$ nodes, a $h$-hopset $H$ of $T$ with $O(n\lambda_h(n))$ edges can be computed in $O(n\lambda_h(n))$ time. A linear size $2(\alpha(n)+1)$-hopset can be computed in $O(n\alpha(n))$ time. In the case $h=3$, the constructed hopset allows to obtain a distance oracle using space of $O(n\log\log n)$ edges of $O(\log n)$ bits and having query time $O(\log^2\log n)$.
\end{proposition}

We note that the  $O(n\lambda_h(n))$ hopset size is indeed tight for some trees.
If $P$ is a path with nodes from $1$ to $n$, any $h$-hopset can be seen as a covering of intervals in $[1,n]$ where $[i,j]$ denotes the interval $i,i+1,\ldots,j$ of integers. More precisely, a set $I$ of intervals $h$-covers $[1,n]$ when every interval $[i,j]\subseteq [1,n]$ is the union of at most $h$ intervals in $I$~\cite{AlonSchieber}. We can easily obtain a $h$-covering from any $h$-hopset $H$ of the path $P$ by associating each edge $uv$ of $P\cup H$ to the interval $[u,v]$. A lower bound of $\Omega(n\lambda_h(n))$ for the size of a $h$-covering of $[1,n]$ is proved in~\cite{AlonSchieber}.

\paragraph{Correctness of construction.}
The correctness of the constructed $h$-hopset $H$ comes from the fact that two nodes $u,v$ in two different connected components of $T\setminus P$ both have an hopset edge to a node in $P$ on the path $P_{uv}$ from $u$ to $v$ according to Lemma~\ref{lem:splittree}. Let $x$ and $y$ denote the nodes in $P\cap P_{uv}$ that are linked to $u$ and $v$ respectively ($ux,vy\in H$). For $h>2$, the $(h-2)$-hopset added in the construction implies that a path of at most $h-2$ hops links $x$ to $y$ in $T\cup H$ and we thus have $d^h_{T\cup H}(u,v)=d_T(u,v)$. For $h=2$, we also have $vx\in H$ (and $uy\in H$), and $x\in P_{uv}$ implies $d^2_{T\cup H}(u,v)=d_T(u,v)$.

\paragraph{Analysis.}
We claim that the resulting $h$-hopset has $O(n\lambda_h(n))$ edges for $h>1$. Recall that a $1$-hopset has $\Theta(n^2)$ edges. Note that the choice of $p=\frac{n}{\lambda_{h-2}(n)}$ in our construction implies that connected components created by Lemma~\ref{lem:splittree} have size at most $\lambda_{h-2}(n)$. The components created in a recursive call with recursion depth $j$ will have size $\lambda_{h-2}^{(j)}(n)$. The number of recursion levels is thus $\min\{ j \mid \lambda_{h-2}^{(j)}(n)\le 1\}=\lambda_h(n)$.
We now show that $O(n)$ edges are added to the construction at each recursion level.
For $h=2$, we have $p=O(1)$ and the number of edges added at each recursion level is thus at most $O(n)$. For $h=3$, we have $p=\frac{n}{\lambda_1(n)}=\sqrt{n}$ and the $(h-2)$-hopset constructed on at most $2p$ nodes has $O(n)$ edges. For $h>3$, we proceed by induction on $h$: we assume that the $(h-2)$-hopset constructed for a tree with at most $2p$ nodes has $O(2p\lambda_{h-2}(2p))$ edges that is $O(n)$ edges for $p=\frac{n}{\lambda_{h-2}(n)}$ (note that $\lambda_{h-2}$ is non-decreasing for any $h>0$).

\paragraph{Query time for $3$-hopsets.}
For the special case of $h=3$, we have $\lambda_3(n) = \log \log n$, and the size required to represent the $3$-hop data structure is $\S = O(n \log \log n)$ edges. Moreover, following the convention $H = H_1 \cup H_2$ introduced in the Preliminaries, we note that in the adopted construction, $\deg_{H_1}(v) = O(\log \log n)$ for all $v \in V$. A bound of $O(\log^2\log n)$ query time follows from the above analysis. 

\paragraph{Linear size hopset.}
We can obtain a linear size $2(\alpha(n)+1)$-hopset by splitting $T$ into sub-trees of size at most $\alpha(n)$ using Lemma~\ref{lem:splittree} with $p=\frac{n}{\alpha(n)}$. Two nodes in a connected component of $T\setminus P$ are thus obviously linked by a path of length at most $\alpha(n)$ in $T$. Similarly as before, we link every node to the (at most 2) nodes in $P$ connected to its component and add a $2\alpha(n)$-hopset for the forest induced by nodes in $P$. We thus obtain a $2(\alpha(n)+1)$-hopset with $O(n + \frac{n}{\alpha(n)}{\lambda_{2\alpha(n)}(n)})=O(n)$ edges.

We now consider the case of bounded treewidth graphs.
\paragraph{$h$-hopset construction for bounded treewidth graphs.}
Consider a graph $G$ with treewidth $t$ and an associated tree $T$.
The general idea is to follow the construction of a $h$-hopset of $T$ with slight modifications. Similarly to the tree case, we select a set $P$ of at most $2p$ bags with $p=\frac{n/t}{\lambda_{h-2}(n/t)}$ according to Lemma~\ref{lem:splittree}. We then construct a $(h-2)$-hopset $H_{T'}$ of the forest $T'$ induced by bags in $P$ according to the tree construction. For each edge $XY$ in $H_{T'}$, we add an edge $xy$ to the graph hopset for all $x\in X$ and $y\in Y$. Such edges are called \emph{tree-hopset} edges.
Now for each node $u$, such that its root bag $R_u$ falls in a connected component of $T\setminus P$, we consider the (at most 2) bags $Y\in P$ that are connected to that component and add an edge $uy$ to the graph hopset for all $y\in Y$. Such edges are called \emph{separator} edges. 
We then recurse on each component of $T\setminus P$ until we reach subtrees of size $n'\le t$. We then pursue with $p=\frac{n'}{\lambda_{h-2}(n')}$ and so on recursively until reaching components of size at most 1.
Finally, for each node $u$, we add an edge $ux$ to the graph hopset for all $x\in R_u$. Such edges are called \emph{bag} edges.
To construct a linear size hopset, we use a single step with $p=\frac{n}{\alpha(n)}$ and a $2\alpha(n)$-hopset of $T'$. For each tree edge $XY$ inside components of $T\setminus P$ we add an edge $xy$ to the construction for all $x\in X$ and $y\in Y$ such that $x\notin Y$ and $y\notin X$. Such edges are also considered as tree-hopset edges. We now state the parameters of the designed hopset.

\begin{theorem}\label{thm:btw}
\label{th:treewidth}
For all $h>1$, any graph with treewidth $t$ has a $h$-hopset with $O(tn\lambda_h(n))$ edges and a $2(\alpha(n)+1)$-hopset with $O(t^2n)$ edges.
\end{theorem}

\paragraph{Correctness of construction.}
Let $H$ denote the hopset constructed for a graph $G$ with treewidth $t$ and associated tree $T$.
Consider a shortest path $Q=u_0,\ldots,u_k$ for some integer $k\ge 1$.
First consider the case where a bag $X$ of $T$ contains both $u_0$ and $u_k$. Without loss of generality, we can assume than $R_{u_0}$ is an ancestor of $R_{u_k}$. As $R_{u_k}$ lies on the path from $X$ to $R_{u_0}$, it must contain $u_0$ and edge $u_ku_0$ is in $H$ according to the last step of the above construction. Now suppose that no bag contains both $u_0$ and $u_k$. Consider the first recursion call where splitting a subtree with a set $P$ of bags separates $R_{u_0}$ and $R_{u_k}$. Consider the path from $R_{u_0}$ to $R_{u_k}$ in $T$. Let $X$ (resp. $Y$) be the first (resp. last) bag in $P$ on that path.
Either $u_0$ is in $X$ or $H$ contains separator edges from $u_0$ to all nodes in $X$. Similarly, either $u_k$ is in $Y$ or $H$ contains a separator edge from $u_k$ to all nodes in $Y$. The $(h-2)$-hopset considered during that recursion call contains a path $P'$ of $h'\le h-2$ hops from $X$ to $Y$. If two consecutive bags contain $u_0$ and $u_k$ respectively, then $H$ contains edge $u_0u_k$ as a tree-hopset edge. Otherwise, let $X'$ (resp. Y') be the first bag in $P'$ not containing $u_0$ (resp. $u_k$). By treewidth definition, there exists bags $X_1,\ldots,X_{k}\in V(T)$ containing edges $u_0u_1,\ldots,u_{k-1}u_k$ respectively (i.e., $X_i$ contains $u_{i-1}$ and $u_i$ for all $i\in\{1,\ldots,k\}$). The shortest path $Q$ corresponds to a walk in $T$ from $X_1$ to $X_2$, then to $X_3$ and so on. All bags on the path (in $T$) from $X_i$ to $X_{i+1}$ must contain $u_i$. As that walk must go through $X'$, we can define the highest index $i_0>0$ such that $u_{i_0}\in X'$. Similarly, we can define the smallest index $j_{0} >= i_0$ such that $u_{j_0}\in Y'$. Our construction $H$ then contains separator edges $u_0u_{i_0}$ and $u_ku_{j_{0}}$.
When $i_0=j_0$, $H$ contains a path of at most 2 hops with same length as $Q$. 
If two consecutive bags of $P'$ contain $u_{i_0}$ and $u_{j_{0}}$ respectively, then $H$ contains a tree-hopset edge $u_{i_0}u_{j_{0}}$. Otherwise, we can similarly define indexes $i_1,\ldots, i_{h''}$ and $j_1,\ldots, j_{h'''}$ with $i_0<i_1<\cdots<i_{h''}<j_{h'''}<\cdots<j_1<j_0$ and $h''+h'''+1\le h'\le h-2$. $H$ then contains tree-hopset edges $u_{i_{0}}u_{i_{1}},\ldots u_{i_{h''-1}}u_{i_{h''}}, u_{i_{h''}}u_{j_{h'''}}, u_{j_{h'''-1}}u_{j_{h'''}}, \ldots, u_{j_{0}}u_{j_{1}}$. Thus, in all cases, $H$ contains a path of at most $h$ hops and same length as $Q$.

\paragraph{Analysis.}
In the first recursion levels, a subtree of size $n'$ is split into subtrees smaller than $t\lambda_{h-2}(n'/t)\le t\lambda_{h-2}(n')$. At recursion depth $\lambda_{h-2}^*(n)=\lambda_h(n)$, we thus obtain subtrees of size at most $t$.
Deeper recursion calls are similar to the tree case. The total number of recursion levels is thus $\lambda_h(n)+\lambda_h(t)=O(\lambda_h(n))$. When processing a subtree of size $n'$, we build a $(h-2)$-hopset for a forest of at most $2p$ bags using $O(2p\lambda_{h-2}(2p))$ edges according to Proposition~\ref{prop:tree}. For $n'>t$, we use $p=\frac{n'/t}{\lambda_{h-2}(n'/t)}$ and thus produce at most $O(t^2\frac{n'}{t})=O(tn)$ tree-hopset edges. For $n'\le t$, we use $p=\frac{n'}{\lambda_{h-2}(n')}$. However, for a given bag $X$, there are at most $n'\le t$ nodes not in $X$ among the other $n'-1$ bags. We thus produce at most $t$ tree-hopset edges per bag. In both cases, each recursion level thus brings $O(tn)$ tree-hopset edges as well as $O(tn)$ separator edges. There are at most $tn$ bag edges in total thus we can obtain a $h$-hopset with $O(t n\lambda_h(n))$ edges for any graph of treewidth $t$.
In the linear size construction, we use a single step using a $2\alpha(n)$-hopset for $T'$ with $O(\frac{n}{\alpha(n)}{\lambda_{2\alpha(n)}(n)})=O(n)$ edges. We thus have $O(t^2n+tn)$ tree-hopset edges and $O(tn)$ separator edges.

\paragraph{Query time for $3$-hopsets.}
For the special case of $h=3$, we have $\lambda_3(n) = \log \log n$, and the size required to represent the $3$-hop data structure is $\S = O(t n \log \log n)$ edges. Following the convention $H = H_1 \cup H_2$, we classify tree-hopset edges in $H_2$ while both separator edges and bag edges are classified in $H_1$. More precisely, when adding separator (or bag) edges $ux$ for all $x$ in a bag $X$, we add $x$ as out-neighbor of $u$ in $H_1$. For any $v\in V$, we thus have $\deg_{H_1}(v)=O(t \log \log n)$. The following bound on the query time follows.

\begin{theorem}
\label{th:treewidth3hop}
Any graph with treewidth $t$ admits a $3$-hopset distance oracle represented on $O(tn \log\log n)$ edges of $O(\log n)$ bits, with a query time of $O(t^2 \log^2\log n)$.
\end{theorem}

\bibliography{main}

\begin{thebibliography}{10}

\bibitem{ABP17}
Amir Abboud, Greg Bodwin, and Seth Pettie.
\newblock A hierarchy of lower bounds for sublinear additive spanners.
\newblock In Philip~N. Klein, editor, {\em Proceedings of the Twenty-Eighth
  Annual {ACM-SIAM} Symposium on Discrete Algorithms, {SODA} 2017, Barcelona,
  Spain, Hotel Porta Fira, January 16-19}, pages 568--576. {SIAM}, 2017.

\bibitem{hwVC}
Ittai Abraham, Daniel Delling, Amos Fiat, Andrew~V. Goldberg, and Renato~F.
  Werneck.
\newblock {VC}-dimension and shortest path algorithms.
\newblock In {\em {ICALP}}, volume 6755 of {\em Lecture Notes in Computer
  Science}, pages 690--699. Springer, 2011.

\bibitem{DBLP:journals/jacm/AbrahamDFGW16}
Ittai Abraham, Daniel Delling, Amos Fiat, Andrew~V. Goldberg, and Renato~F.
  Werneck.
\newblock Highway dimension and provably efficient shortest path algorithms.
\newblock {\em J. {ACM}}, 63(5):41:1--41:26, 2016.

\bibitem{MS-TR}
Ittai Abraham, Daniel Delling, Amos Fiat, Andrew~V. Goldberg, and Renato~F.
  Werneck.
\newblock Highway dimension and provably efficient shortest path algorithms.
\newblock {\em J. {ACM}}, 63(5):41:1--41:26, 2016.

\bibitem{hw10}
Ittai Abraham, Amos Fiat, Andrew~V. Goldberg, and Renato~F. Werneck.
\newblock Highway dimension, shortest paths, and provably efficient algorithms.
\newblock In Moses Charikar, editor, {\em Proceedings of the Twenty-First
  Annual {ACM-SIAM} Symposium on Discrete Algorithms, {SODA} 2010, Austin,
  Texas, USA, January 17-19, 2010}, pages 782--793. {SIAM}, 2010.

\bibitem{DBLP:conf/www/AkibaIY14}
Takuya Akiba, Yoichi Iwata, and Yuichi Yoshida.
\newblock Dynamic and historical shortest-path distance queries on large
  evolving networks by pruned landmark labeling.
\newblock In Chin{-}Wan Chung, Andrei~Z. Broder, Kyuseok Shim, and Torsten
  Suel, editors, {\em 23rd International World Wide Web Conference, {WWW} '14,
  Seoul, Republic of Korea, April 7-11, 2014}, pages 237--248. {ACM}, 2014.

\bibitem{AlonSchieber}
Noga Alon and Baruch Schieber.
\newblock Optimal preprocessing for answering on-line product queries.
\newblock Technical Report 71/87, Tel Aviv University, 1987.

\bibitem{Sublinear}
Stephen Alstrup, S{\o}ren Dahlgaard, Mathias B{\ae}k~Tejs Knudsen, and Ely
  Porat.
\newblock Sublinear distance labeling.
\newblock In Piotr Sankowski and Christos~D. Zaroliagis, editors, {\em 24th
  Annual European Symposium on Algorithms, {ESA} 2016, August 22-24, 2016,
  Aarhus, Denmark}, volume~57 of {\em LIPIcs}, pages 5:1--5:15. Schloss
  Dagstuhl - Leibniz-Zentrum fuer Informatik, 2016.

\bibitem{AngelidakisMO17}
Haris Angelidakis, Yury Makarychev, and Vsevolod Oparin.
\newblock Algorithmic and hardness results for the hub labeling problem.
\newblock In Philip~N. Klein, editor, {\em Proceedings of the Twenty-Eighth
  Annual {ACM-SIAM} Symposium on Discrete Algorithms, {SODA} 2017, Barcelona,
  Spain, Hotel Porta Fira, January 16-19}, pages 1442--1461. {SIAM}, 2017.

\bibitem{Arikati96}
Srinivasa~Rao Arikati, Danny~Z. Chen, L.~Paul Chew, Gautam Das, Michiel H.~M.
  Smid, and Christos~D. Zaroliagis.
\newblock Planar spanners and approximate shortest path queries among obstacles
  in the plane.
\newblock In Josep D{\'{\i}}az and Maria~J. Serna, editors, {\em Algorithms -
  {ESA} '96, Fourth Annual European Symposium, Barcelona, Spain, September
  25-27, 1996, Proceedings}, volume 1136 of {\em Lecture Notes in Computer
  Science}, pages 514--528. Springer, 1996.

\bibitem{DBLP:conf/dimacs/BastFM06}
H.~Bast, Stefan Funke, and Domagoj Matijevic.
\newblock Ultrafast shortest-path queries via transit nodes.
\newblock In Camil Demetrescu, Andrew~V. Goldberg, and David~S. Johnson,
  editors, {\em The Shortest Path Problem, Proceedings of a {DIMACS} Workshop,
  Piscataway, New Jersey, USA, November 13-14, 2006}, volume~74 of {\em
  {DIMACS} Series in Discrete Mathematics and Theoretical Computer Science},
  pages 175--192. {DIMACS/AMS}, 2006.

\bibitem{BhattacharyyaGJRW12}
Arnab Bhattacharyya, Elena Grigorescu, Kyomin Jung, Sofya Raskhodnikova, and
  David~P. Woodruff.
\newblock Transitive-closure spanners.
\newblock {\em {SIAM} J. Comput.}, 41(6):1380--1425, 2012.

\bibitem{BodlaenderTS94}
Hans~L. Bodlaender, Gerard Tel, and Nicola Santoro.
\newblock Trade-offs in non-reversing diameter.
\newblock {\em Nord. J. Comput.}, 1(1):111--134, 1994.

\bibitem{Cabello12}
Sergio Cabello.
\newblock Many distances in planar graphs.
\newblock {\em Algorithmica}, 62(1-2):361--381, 2012.

\bibitem{ChaudhuriZ00}
Shiva Chaudhuri and Christos~D. Zaroliagis.
\newblock Shortest paths in digraphs of small treewidth. part {I:} sequential
  algorithms.
\newblock {\em Algorithmica}, 27(3):212--226, 2000.

\bibitem{Chazelle87a}
Bernard Chazelle.
\newblock Computing on a free tree via complexity-preserving mappings.
\newblock {\em Algorithmica}, 2:337--361, 1987.

\bibitem{Xu2000}
Danny~Z. Chen and Jinhui Xu.
\newblock Shortest path queries in planar graphs.
\newblock In {\em Proceedings of the Thirty-Second Annual {ACM} Symposium on
  Theory of Computing, May 21-23, 2000, Portland, OR, {USA}}, pages 469--478,
  2000.

\bibitem{Cohen97}
Edith Cohen.
\newblock Using selective path-doubling for parallel shortest-path
  computations.
\newblock {\em J. Algorithms}, 22(1):30--56, 1997.

\bibitem{Cohen00}
Edith Cohen.
\newblock Polylog-time and near-linear work approximation scheme for undirected
  shortest paths.
\newblock {\em J. {ACM}}, 47(1):132--166, 2000.

\bibitem{Cohen:2003:RDQ:942270.944300}
Edith Cohen, Eran Halperin, Haim Kaplan, and Uri Zwick.
\newblock Reachability and distance queries via 2-hop labels.
\newblock {\em SIAM J. Comput.}, 32(5):1338--1355, May 2003.

\bibitem{Cohen-AddadDW17}
Vincent Cohen{-}Addad, S{\o}ren Dahlgaard, and Christian Wulff{-}Nilsen.
\newblock Fast and compact exact distance oracle for planar graphs.
\newblock In {\em 58th {IEEE} Annual Symposium on Foundations of Computer
  Science, {FOCS} 2017, Berkeley, CA, USA, October 15-17, 2017}, pages
  962--973, 2017.

\bibitem{DBLP:conf/esa/DellingGPW14}
Daniel Delling, Andrew~V. Goldberg, Thomas Pajor, and Renato~F. Werneck.
\newblock Robust distance queries on massive networks.
\newblock In Andreas~S. Schulz and Dorothea Wagner, editors, {\em Algorithms -
  {ESA} 2014 - 22th Annual European Symposium, Wroclaw, Poland, September 8-10,
  2014. Proceedings}, volume 8737 of {\em Lecture Notes in Computer Science},
  pages 321--333. Springer, 2014.

\bibitem{Djidjev96}
Hristo Djidjev.
\newblock On-line algorithms for shortest path problems on planar digraphs.
\newblock In Fabrizio d'Amore, Paolo~Giulio Franciosa, and Alberto
  Marchetti{-}Spaccamela, editors, {\em Graph-Theoretic Concepts in Computer
  Science, 22nd International Workshop, {WG} '96, Cadenabbia (Como), Italy,
  June 12-14, 1996, Proceedings}, volume 1197 of {\em Lecture Notes in Computer
  Science}, pages 151--165. Springer, 1996.

\bibitem{ElkinN16}
Michael Elkin and Ofer Neiman.
\newblock Hopsets with constant hopbound, and applications to approximate
  shortest paths.
\newblock In Irit Dinur, editor, {\em {IEEE} 57th Annual Symposium on
  Foundations of Computer Science, {FOCS} 2016, 9-11 October 2016, Hyatt
  Regency, New Brunswick, New Jersey, {USA}}, pages 128--137. {IEEE} Computer
  Society, 2016.

\bibitem{ElkinN17}
Michael Elkin and Ofer Neiman.
\newblock Linear-size hopsets with small hopbound, and distributed routing with
  low memory.
\newblock {\em CoRR}, abs/1704.08468, 2017.

\bibitem{FR06}
Jittat Fakcharoenphol and Satish Rao.
\newblock Planar graphs, negative weight edges, shortest paths, and near linear
  time.
\newblock {\em J. Comput. Syst. Sci.}, 72(5):868--889, 2006.

\bibitem{FarzanK14}
Arash Farzan and Shahin Kamali.
\newblock Compact navigation and distance oracles for graphs with small
  treewidth.
\newblock {\em Algorithmica}, 69(1):92--116, 2014.

\bibitem{Gavoille:2004:DLG:1036161.1036165}
Cyril Gavoille, David Peleg, St{\'e}phane P{\'e}rennes, and Ran Raz.
\newblock Distance labeling in graphs.
\newblock {\em J. Algorithms}, 53(1):85--112, October 2004.

\bibitem{DBLP:conf/wdag/GawrychowskiKU16}
Pawel Gawrychowski, Adrian Kosowski, and Przemyslaw Uznanski.
\newblock Sublinear-space distance labeling using hubs.
\newblock In {\em Distributed Computing - 30th International Symposium, {DISC}
  2016, Paris, France, September 27-29, 2016. Proceedings}, pages 230--242,
  2016.

\bibitem{GawrychowskiMWW18}
Pawel Gawrychowski, Shay Mozes, Oren Weimann, and Christian Wulff{-}Nilsen.
\newblock Better tradeoffs for exact distance oracles in planar graphs.
\newblock In {\em Proceedings of the Twenty-Ninth Annual {ACM-SIAM} Symposium
  on Discrete Algorithms, {SODA} 2018, New Orleans, LA, USA, January 7-10,
  2018}, pages 515--529, 2018.

\bibitem{DBLP:conf/wea/GeisbergerSSD08}
Robert Geisberger, Peter Sanders, Dominik Schultes, and Daniel Delling.
\newblock Contraction hierarchies: Faster and simpler hierarchical routing in
  road networks.
\newblock In Catherine~C. McGeoch, editor, {\em Experimental Algorithms, 7th
  International Workshop, {WEA} 2008, Provincetown, MA, USA, May 30-June 1,
  2008, Proceedings}, volume 5038 of {\em Lecture Notes in Computer Science},
  pages 319--333. Springer, 2008.

\bibitem{RE}
Andrew~V. Goldberg, Haim Kaplan, and Renato~F. Werneck.
\newblock Reach for {A}*: Efficient point-to-point shortest path algorithms.
\newblock In {\em {ALENEX}}, pages 129--143. {SIAM}, 2006.

\bibitem{reach}
Ronald~J. Gutman.
\newblock Reach-based routing: {A} new approach to shortest path algorithms
  optimized for road networks.
\newblock In {\em {ALENEX/ANALCO}}, pages 100--111. {SIAM}, 2004.

\bibitem{KS97}
Philip~N. Klein and Sairam Subramanian.
\newblock A randomized parallel algorithm for single-source shortest paths.
\newblock {\em J. Algorithms}, 25(2):205--220, 1997.

\bibitem{KV-long2016}
Adrian Kosowski and Laurent Viennot.
\newblock Beyond highway dimension: Small distance labels using tree skeletons.
\newblock {\em CoRR}, abs/1609.00512, 2016.

\bibitem{KV-SODA2017}
Adrian Kosowski and Laurent Viennot.
\newblock Beyond highway dimension: Small distance labels using tree skeletons.
\newblock In Philip~N. Klein, editor, {\em Proceedings of the Twenty-Eighth
  Annual {ACM-SIAM} Symposium on Discrete Algorithms, {SODA} 2017, Barcelona,
  Spain, Hotel Porta Fira, January 16-19}, pages 1462--1478. {SIAM}, 2017.

\bibitem{PelegS89}
David Peleg and Alejandro~A. Sch{\"{a}}ffer.
\newblock Graph spanners.
\newblock {\em Journal of Graph Theory}, 13(1):99--116, 1989.

\bibitem{RaghavanT87}
Prabhakar Raghavan and Clark~D. Thompson.
\newblock Randomized rounding: a technique for provably good algorithms and
  algorithmic proofs.
\newblock {\em Combinatorica}, 7(4):365--374, 1987.

\bibitem{SS99}
Hanmao Shi and Thomas~H. Spencer.
\newblock Time-work tradeoffs of the single-source shortest paths problem.
\newblock {\em J. Algorithms}, 30(1):19--32, 1999.

\bibitem{Thorup95}
Mikkel Thorup.
\newblock Shortcutting planar digraphs.
\newblock {\em Combinatorics, Probability {\&} Computing}, 4:287--315, 1995.

\bibitem{Thorup97}
Mikkel Thorup.
\newblock Parallel shortcutting of rooted trees.
\newblock {\em J. Algorithms}, 23(1):139--159, 1997.

\bibitem{UY91}
Jeffrey~D. Ullman and Mihalis Yannakakis.
\newblock High-probability parallel transitive closure algorithms.
\newblock In {\em {SPAA}}, pages 200--209, 1990.

\end{thebibliography}

\end{document}